\documentclass[12pt]{article}
\usepackage{epsfig}
\usepackage{graphicx}
\usepackage{amssymb,amsmath}
\usepackage{hyperref}
\usepackage[nosort]{cite}
\usepackage{booktabs}
\usepackage{multirow}
\usepackage{caption}
\usepackage{comment}
\usepackage{subcaption}

\hypersetup{
    colorlinks=true,
    linkcolor=blue,
    citecolor=red,
}

\usepackage{xcolor}

\def\dd{d\!\!{}^-\!}
\def\del{\delta\!\!\!{}^-\!}

\begin{document}

\begin{center}
\Large{\bf Hyper-Hermitian Weyl Double Copy} \vspace{0.5cm}

\large  E. Chac\'on\footnote{e-mail address: {\tt
erick.cr@lapaz.tecnm.mx}}\textsuperscript{,†} , H. Garc\'{\i}a-Compe\'an\footnote{e-mail address: {\tt
hugo.compean@cinvestav.mx}}\textsuperscript{,§}, G. Robles\footnote{e-mail
address: {\tt gilberto.robles@cinvestav.mx}}\textsuperscript{,§}

\vspace{0.3cm}

\textsuperscript{†}{\small \em Departamento de Ciencias Básicas, Tecnológico Nacional de México \\ CP 23080, La Paz, B.C.S., México}\\

\vspace{0.3cm}

\textsuperscript{§}{\small \em Departamento de F\'{\i}sica, Centro de
Investigaci\'on y de Estudios Avanzados del IPN}\\
{\small\em P.O. Box 14-740, CP. 07000, Ciudad de M\'exico, Mexico}\\

\vspace*{1.5cm}
\end{center}

\begin{abstract}
The self-dual double copy is further explored. In previous work, it has been shown that hyper-Hermitian manifolds 
also have associated the self-dual gauge theories via Kerr-Schild double copy. The self-dual double copy is generalized in the structure of the kinematic algebra by replacing the area-preserving diffeomorphisms algebra with the diffeomorphisms on a surface algebra. This gave rise to the hyper-Hermitian double copy in the Kerr-Schild approach. In the present article, we further study the hyper-Hermitian case using the Weyl double copy formalism. In particular, we have found solutions within this formalism for different hyper-Hermitian metrics. One of the main features is that there will be two Maxwell spinors and one of them is source-free while the other has a source current. This is compatible with the fact that, in general, the hyper-Hermitian spaces are not Ricci-flat.

\vskip 1truecm

\end{abstract}

\bigskip

\newpage

\tableofcontents

\newpage

\section{Introduction}
\label{S-Intro}

In the past, the interplay between different theoretical physics subjects has been a way of
understanding physical phenomena and is a road to advancing in physics. For instance, the AdS/CFT \- \- \- (or gauge/gravity) co\-rres\-pondence is a duality which (in its more traditional version) connects strong coupling supersymmetric quantum conformal gauge theory in dimension $d$ with
classical gravity theory in $d+1$ dimensions (see for instance, ref. \cite{Aharony:1999ti}). Another relation between gauge and gravitational theories is the one of the double copy and the Bern-Carrasco-Johansson (BCJ) duality \cite{Bern:2008qj,Bern:2010yg,Bern:2010ue} which relates scattering amplitudes of non-abelian gauge theories in $d$ dimensions and scattering amplitudes of gravitational theories in the same dimension $d$ (for a complete review, see ref. \cite{Bern:2019prr,Adamo:2022dcm}). The tree-level version has been proved and the loop extension is still at the level of conjecture.

The ideas of the double copy and the BCJ duality have been applied at the level of equations of motion in gauge and gravity theories. Thus it is promoted to the level of solutions of gauge and gravity theories. There is a correspondence between solutions of the equation of motion in both theories \cite{Monteiro:2014cda}. The original version of the classical double copy uses the Kerr-Schild ansatz which implies the existence of certain coordinate systems in which the  metric of a gravity theory can be expressed as $g_{\mu \nu} = \eta_{\mu \nu} + \phi k_\mu k_\nu$.  Many applications in diverse field theories were studied for this classical double copy approach, for instance, see refs. \cite{Luna:2015paa,Luna:2016due,Ridgway:2015fdl,Bahjat-Abbas:2017htu,Luna:2016hge,Carrillo-Gonzalez:2017iyj,CarrilloGonzalez:2019gof,Berman:2018hwd,Bah:2019sda,Keeler:2020rcv,Easson:2020esh,Easson:2021asd,Easson:2022zoh,Alkac:2021seh,Gonzo:2021drq,Armstrong-Williams:2022apo, Didenko:2022qxq}. Later another possibility to study the correspondence was through the use of Weyl spinors, which was named the Weyl double copy \cite{Luna:2018dpt}. The Weyl double copy is reduced to the  Kerr-Schild double copy in type {\bf D} spacetimes \cite{Easson:2023dbk}. The Weyl double copy relates the curvature of gravity theory written in terms of Weyl spinors to the  field strength spinor of a gauge field. After its proposal, Weyl double copy was discussed in different contexts in refs. \cite{Sabharwal:2019ngs,Alawadhi:2020jrv,Godazgar:2020zbv,Alkac:2023glx,Armstrong-Williams:2024bog}.
This was the precursor to finding a more general description of the double copy in terms of twistor theory  \cite{White:2020sfn,Chacon:2021wbr,Chacon:2021hfe,Chacon:2021lox,Luna:2022dxo,Borsten:2023paw,Beetar:2024ptv}. One of the successes of the double copy has been its application as an efficient tool for computing processes involving gravitational waves  \cite{Godazgar:2020zbv}.

The case for the self-dual sector of the Yang-Mills theories in four dimensions and its correspondence with the self-dual sector of Einstein's theory was proposed and explored for the first time within the context of the double copy\cite{Monteiro:2011pc,Adamo:2022lah,Monteiro:2013rya}. Integrability anomalies arising in the self-dual sectors
of gravity and Yang-Mills theory are studied in ref. \cite{Doran:2023cmj}. This case is of particular relevance since both gauge and gravitational sectors can be described through a cubic Lagrangian giving rise to trivalent Feynman graphs. In the self-dual gauge case, one has the Leznov-Parks Lagrangian \cite{Leznov:1986up,Parkes:1992rz} and in the gravitational self-dual sector, we have the Pleba\'nski Lagrangian \cite{Boyer:1979hc,Plebanski:1996np}. In ref. \cite{Chacon:2020fmr}  new cases for the heavenly double copies were studied. There, integrable and non-integrable deformations of the self-dual double copy and generalizations of it as the hyper-Hermitian case were considered. The double copy has been studied as well in the context of celestial amplitudes \cite{Casali:2020vuy,Monteiro:2022lwm,Monteiro:2022xwq} and higher spin self-dual theories \cite{Ponomarev:2017nrr}. Some recent works regarding further developments of the self-dual double copy were discussed in refs. \cite{Brown:2023zxm,CarrilloGonzalez:2024sto,LopesCardoso:2024ttc,Miller:2024oza}.

In the present article, we continue the study, which started in ref. \cite{Chacon:2020fmr} and we carry out a deeper study of the hyper-Hermitian double copy using the Weyl double copy.  The hyper-Hermitian metrics have been studied previously in refs. \cite{CBoyer:1988,Chave:1991pv,Gauduchon:1998,Grant:1998qlz,Dunajski:1998nj}. In ref. \cite{Dunajski:1998nj}, hyper-Hermitian complex metrics are determined through two holomorphic functions $\Theta_A(x^a)$ (with $A=0,1$) of the coordinates on a complex four manifold. Moreover $\Theta_A(x^a)$ satisfies a generalization of the Pleba\'nski's second heavenly equation \cite{Plebanski:1975wn} that we term it  {\it Dunajski's heavenly equation}. Indeed, this generalized equation can be reduced to the second heavenly equation under the implementation of a condition on $\Theta_A(x^a)$. For a derivation of the hyper-Hermitian metric satisfying a generalization of the first heavenly equation, see ref. \cite{schucking}.   In the present article, we carry out a systematic study of the Weyl double copy for the hyper-Hermitian metrics, focusing on the family of solutions found in ref. \cite{Dunajski:1998nj}. One of the main results of our work is that, for the mentioned family of solutions, the Weyl double copy correspondence implies that one of the two Maxwell  fields  has a source current while the other remains sourceless. Such a current is, in general, non-conserved.

The paper is organized as follows: in Section 2 we give a brief overview of the double copy for the self-dual case. We describe first the self-dual sector with the kinematic factor taking values in the Lie algebra of area-preserving diffeomorphisms and its eventual generalization to the kinematic algebra consisting of the Lie algebra of smooth vector fields for the hyper-Hermitian case.  In Section 3, after a brief overview on the Weyl double copy, we carry out a systematic study of the Weyl double copy for the hyper-Hermitian metrics. We will focus on the family of metrics called elementary states introduced in ref. \cite{Dunajski:1998nj}. In Section 4 we provide explicit examples of solutions of the hyper-Hermitian double copy in the language of Weyl spinors. Finally, Section 5 is devoted to give our final remarks. 

\section{Brief Overview of the Classical Double Copy}
\label{SDDDC}
In the present section, we briefly overview the double copy procedure between the self-dual sector of the four-dimensional Yang-Mills theory  and self-dual gravity in Euclidean signature. Our aim is not to provide an extensive description but to serve to introduce notation and conventions that we will follow in the subsequent sections.  

We use local coordinates $x^a=\{x,y,u,v\}$\footnote{In ref. \cite{Chacon:2020fmr} the subject was discussed in the Lorentz signature: $(+,-,-,-)$. There the local coordinates were $\{\widetilde{u},\widetilde{v},W,\overline{W}\}$. The relation between these coordinates and those used in the present article in the Euclidean formulation are: $\widetilde{u} =-iy$, $\widetilde{v}= iv$, $W=-ix$, $\overline{W}=-iu$. In terms of these coordinates, the Lorentzian metric is given by: $ds^2_L = d\widetilde{u} d\widetilde{v} - dW d\overline{W}.$}. The Euclidean metric in these coordinates reads: $ds^2_E= dx du + dy dv$. 
\subsection{Hyper-K\"ahler double copy}
\label{sec_selfdualDC}

It is known that the double copy has a realization in terms of the equation of motion of self-dual Yang-Mills theory starting from a Lagrangian with cubic interactions \cite{Monteiro:2011pc,Adamo:2022lah}. In this case, the kinematic algebra is known to be precisely the area-preserving diffeomorphism algebra $\mathfrak{sdiff}$. The Lagrangian formulation used to show this explicitly was based on the description of Leznov and Parkes in which the dynamics of the self-dual sector of Yang-Mills theory is governed by a Lagrangian for a scalar field with a cubic interaction term \cite{Leznov:1986up,Parkes:1992rz}, whose action is given in Euclidean signature by
\begin{equation}
I= \int d^4x {\rm Tr} \bigg\{ - {1 \over 2} \big[(\partial_x \phi) (\partial_u \phi) + (\partial_y \phi) (\partial_v \phi)\big]+ {g\over 3} \phi [\partial_x \phi, \partial_y \phi] \bigg\} 
\end{equation}
for a Lie algebra-valued scalar field $\phi(x) =\phi^a(x)T_a$ on a $4$-dimensional space, where $T_a$ are anti-hermitian matrix elements of the Lie algebra
$\mathfrak{g}$ and $a=1,\cdots, {\rm dim} \ \mathfrak{g}$. The equations of motion of this action are written as
\begin{equation}
\partial^2\phi-ig[\partial_x\phi,\partial_y\phi]=0,
\label{YMEOM}
\end{equation}
where $g$ is the Yang-Mills coupling constant, $\partial^2 \equiv \partial_x \partial_{u} + \partial_y \partial_{v}$ and  $\phi$ is a function of the local coordinates on $\mathbb{R}^4$. In the momentum-space, through the use of the Fourier transform, we have
\begin{equation}
\phi(k)=-ig\int \dd p_1\dd p_2\frac{\del(p_1+p_2-k)}{k^2}
X(p_1,p_2)\phi(p_1)\phi(p_2), 
\label{SDYMFourier}
\end{equation}
where $k^2 \equiv k_x k_{u} + k_y k_{v}$. In the previous equation we are following the definitions $\dd p\equiv \frac{d^4 p}{(2\pi)^4},$ $\del(p)\equiv(2\pi)^4\delta^{(4)}(p)$ and, $X(p_1,p_2)\equiv p_{1y}p_{2x}-p_{1x}p_{2y}$ is the kinematic structure constant. Expression (\ref{SDYMFourier}) can be recast in terms of $\phi^a(x)$ in the following form
\begin{equation}
\phi^a(k)=\frac{g}{2}\int\dd{p}_1\,\dd{p}_2\,\frac{\del(p_1+p_2-k)}{k^2}
\,X(p_1,p_2)\,f^{a b c}\,\phi^{b}(p_1)\phi^{c}(p_2).
\label{YMsol}
\end{equation}
where $f^{abc}$ are the structure constants of
$\mathfrak{g}$ (in some representation of the gauge group), which fulfils $[T^{a},T^{b}]=if^{abc}T^{c}$. As part of the BCJ duality, it is remarkable to see that there is a choice of the spacetime coordinates where the kinematic structure constants $X(p_1,p_2)$ also fulfil a Jacobi identity just as the gauge structure constants do. Thus, from the structure of eq. (\ref{YMsol}) one can see the Lie algebra structure $\mathfrak{sdiff} \times\mathfrak{g}.$ 

In the double copy for the classical equations of motion, the second heavenly equation of Pleba\'nski is favoured with respect to the first one. As we reviewed for the Yang-Mills case, eq. (\ref{YMsol}), there are kinematic factors $X(p_1,p_2)$ which satisfies the mentioned kinematic algebra $\mathfrak{sdiff}$ and color factors $f^{abc}$. The double copy correspondence interchanges the color factors with another kinematic factor $f^{abc}\leftrightarrow X(p_1,p_2)$ and the gauge and gravity coupling constants $g \leftrightarrow \kappa$. Thus, we obtain the gravity equation of motion in momentum space
\begin{equation}
\theta(k)=\frac{\kappa}{2}\int\dd p_1\dd p_2\,\frac{\del(p_1+p_2-k)}{k^2}
\,X(p_1,p_2) X(p_1,p_2)\,\theta(p_1)\theta(p_2),
\label{Gravsol} 
\end{equation}
for a scalar field $\theta$. We now identify the coupling constant as the gravitational coupling $\kappa=\sqrt{32\pi G_N}$, where $G_N$ is Newton's constant. In coordinate space, eq.~(\ref{Gravsol}) is simply the {\it Pleba\'nski equation } for self-dual gravity, also known as the  (second) {\it heavenly equation} \cite{Plebanski:1975wn}:
\begin{equation}
\partial^2\theta+\kappa\{\partial_x\theta,\partial_{y}\theta\}=0,
\label{Plebanski}
\end{equation}
where the Poisson bracket in eq. (\ref{Plebanski}) is given with respect to the variables $x$ and $y$. Thus, the double copy in the self-dual sector of Yang-Mills theory is particularly transparent, and amounts to simply replacing the color algebra with its kinematic counterpart, which is the area-preserving diffeomorphism algebra $\mathfrak{sdiff}$. Thus, the structure is given by $\mathfrak{sdiff} \times \mathfrak{sdiff}.$

Eq.~\eqref{Plebanski} reproduces Plebanski's second heavenly equation in Euclidean signature \cite{Plebanski:1975wn}:
\begin{equation}
\partial_x \partial_u \theta + \partial_y \partial_v \theta + \partial_x \partial_x \theta \cdot  \partial_y \partial_y \theta - \partial_x \partial_y \theta \cdot \partial_x \partial_y \theta=0,
\label{Plebanski_Des}
\end{equation}
where we have taken $\kappa=1$. The metric associated with eq.~\eqref{Plebanski_Des} is given by \cite{Plebanski:1975wn}
\begin{equation}
d s^2 = dx du + dy dv + (\partial_y \partial_y \theta ) du^2-2 (\partial_y \partial_x \theta) du dv +( \partial_x \partial_x \theta ) dv.
\label{MetricPlebanski} 
\end{equation}
A class of important solutions is obtained when we take both terms equal to zero  in eq.~\eqref{Plebanski}
\begin{equation}
\partial^2\theta = \{\partial_x\theta,\partial_y\theta\}=0.
\label{ElementaryHK}
\end{equation}

A solution of eq.~\eqref{ElementaryHK} is the Sparling-Tod metric \cite{Sparling:1981nk}
\begin{equation}
d{s}^2= d{u}d{x} + d{y}d{v} + 2 \frac{( u d{v}- v d{u})^2}{(u x+y v)^3} ,
\end{equation}
this metric has precisely the form of eq.~\eqref{MetricPlebanski}.

\subsection{Hyper-Hermitian double copy}
\label{sec_hyper-hermitian}

Now we give a brief overview of  hyper-Hermitian spaces (for more details, see refs. \cite{CBoyer:1988,Dunajski:1998nj}). We begin our exploration with a four-dimensional Riemannian manifold denoted as $(\mathcal{M}, g)$, where $g$ represents a metric that can be either Lorentzian or Euclidean on $\mathcal{M}$. The metric $g$ is hyper-Hermitian if it is Hermitian with respect to each one of the 3 complex structures $I$, $J$, and $K$ on the manifolds ${\cal M}$, such that they satisfy the relations:  
$IJ=-JI=K$ plus cyclic permutations of $I$, $J$ and $K$. In this case, the metric $g$ is of the form
\begin{equation}
ds^2=dudx+dvdy-\left[(\partial_{x} \Theta^{0})dv^2-(\partial_{y} \Theta^{0}+\partial_x \Theta^1)dvdu + (\partial_{y} \Theta^{1})du^2\right],
\label{metricHyperHermitian}
\end{equation}
where $\Theta^A = (\Theta^0,\Theta^1)$ is a pair of complex-valued functions (or potentials) on ${\cal M}$ which satisfy the Dunajski's heavenly equations \cite{Dunajski:1998nj}
\begin{equation}
\partial^{2}\Theta^{A}+\{\Theta^{B},\partial_{B}\Theta^{A}\} =0,
\label{HH}
\end{equation}
where  $x^A=\{y,-x\}$ and $\partial_{A} \equiv \partial_{x^{A}}$.  It is shown in \cite{Dunajski:1998nj} that the metric of eq. (\ref{metricHyperHermitian}) is a hyper-Hermitian metric and every hyper-Hermitian metric locally arises under this construction. Within the formalism,  a relation with the hyper-K\"ahler case described by Plebanski's second heavenly eq. (\ref{Plebanski_Des}) is shown. Such a relation is given by the existence of a potential $\Theta^A$ such that $\nabla_{A0'} \Theta^A =0$, which in turn ensures locally the existence of a scalar function $\theta$ that satisfies eq. (\ref{Plebanski_Des}). Every hyper-Hermitian manifold admits a Lee form, which is a one-form $A$ depending on the manifold metric \cite{CBoyer:1988,Dunajski:1998nj}. In this context, the Lee form is given in terms of the potential $\Theta^A$ in the following form
\begin{equation}
A = {\partial^2 \Theta^B \over \partial x^A \partial x^B} du^A,
\label{photon}
\end{equation}
where $u^A=(u,v)$.

Thus, in momentum space, eq. ~(\ref{HH})
implies the integral equations
\begin{equation}
\Theta^{A}(k)=\frac{1}{2} \int \dd p_{1}\dd p_{2}\frac{\del (p_{1}+p_{2}-k)}{k^{2}}X (p_{1},p_{2})Y^{A}(p_{1C},p_{2B})\Theta^{B}(p_{1})\Theta^{C}(p_{2}),
\label{MomentumHH}
\end{equation}
where $Y^{A}(p_{1C},p_{2B}) \equiv p_{2B}\delta^A_C - p_{1C}\delta^A_B$ is a kinematic factor. It can be proved \cite{Chacon:2020fmr} that they satisfy the anti-symmetric property and the Jacobi identity of the Lie algebra of diffeomorphisms of a compact surface $\mathfrak{diff}$. These constitute the Lie algebra of smooth vector fields 
on a two-dimensional manifold. They are not Hamiltonian vector fields, as they do not satisfy the volume-preserving condition as was discussed in ref. \cite{Dunajski:1998nj}. As it was elaborated in ref. \cite{Chacon:2020fmr}, eq. (\ref{MomentumHH}) can be regarded as the double copy of eq. (\ref{YMsol}) via the BCJ mapping $f^{abc}\rightarrow Y^A(p_{1C},p_{2B})$. This leads to a  $\mathfrak{sdiff}\; \times\; \mathfrak{diff}$ theory.

Eq. (\ref{HH}) can be rewritten in coordinates $(x,y,u,v)$ as follows
\begin{equation}
\partial_x \partial_u \Theta^A + \partial_y \partial_v \Theta^A + \partial_x  \Theta^0 \partial_y \partial_y \Theta^A -
 \partial_x  \Theta^1 \partial_y \partial_x \Theta^A -   \partial_y \Theta^0 \partial_x \partial_y \Theta^A
 +  \partial_y  \Theta^1 \partial_x \partial_x \Theta^A=0.
\end{equation}
A class of solutions to eq. (\ref{HH}) is	 called the hyper-Hermitian elementary states, and they satisfy the following equations (which are the hyper-Hermitian analogous of eq. (\ref{ElementaryHK}))
\begin{equation}
\partial^{2}\Theta^{A}=\{\Theta^{B},\partial_{B}\Theta^{A}\} =0.
\label{igualdad}
\end{equation} 
In the subsequent sections, we will study the double-copy correspondence associated with these elementary states, which are of
$$
\Theta_A=\frac{1}{u x + y v} F_A\left( \frac{u}{u x + y v},\frac{v}{u x + y v}\right), 
$$
where $F_A$ is a pair of complex-valued functions with explicit dependence on their arguments.

\section{Hyper-Hermitian and Hyper-K\"ahlerian Weyl Double Copy}
\label{WDC}
In the present section, we study a certain class of hyper-Hermitian metrics called elementary states. After some preliminary material on the Weyl double copy, we introduce  useful definitions. We apply the Weyl double copy to perform a systematic analysis of the double and single copies of these elementary states.

\subsection{Overview of the Weyl double copy}
In the present subsection, we provide the procedure for constructing double-copy examples. In this and subsequent sections, for notation and conventions, we follow the Penrose and Rindler classic book \cite{Penrose:1986ca}. 
Thus, we first recall some general aspects of the Weyl double copy following mainly ref. \cite{Luna:2018dpt}. Consider the Weyl curvature $\Psi_{ABCD}$ with the symmetry property $\Psi_{ABCD} = \Psi_{(ABCD)}$, where $A = (0,1)$ is a  spinorial index. In the Weyl double copy, the Weyl curvature can be written as
\begin{equation}
\label{WDC2}
	\Psi_{ABCD} = \frac{\Phi^{(1)}_{(A B} \Phi^{(2)}_{CD)}}{\Phi}.
\end{equation}
In general, the Maxwell field strength spinors $\Phi^{(1)}{AB}$ and $\Phi^{(2)}{AB}$ are different; in the single copy, there is only one Maxwell spinor. Then, eq.~\eqref{WDC2} expresses the relationship between Einstein and Maxwell solutions. 
We propose a family of $\Phi$ functions that we will use in the rest of the paper which is given by
\begin{equation}
\label{ZeroCopy}
\Phi=A \frac{1}{(u x + yv)}+B_r \frac{ \left(u x+y v\right)^r}{v^2 u^{r-1}} +C_r \frac{ \left(u x+y v\right)^r}{u^2 v^{r-1}},
\end{equation}
for all $r \in \mathbb R$, where $A$, $B_r$ and $C_r$ are constants. Moreover, $\Phi$ is a solution of the wave equation \begin{equation}
\label{eqDeOnda}
\partial^{\mu}\partial_{\mu} \Phi = 0.
\end{equation} 
We note that we can change $u \leftrightarrow v$, $x \leftrightarrow y$ in $\Phi$ and it will still satisfy eq.~\eqref{eqDeOnda}. These types of properties are observed for example in ref. \cite{Brown:2023zxm}.
To obtain the Maxwell tensor field we use the spinorial formalism. We have that the 'spinorial vierbein' $\sigma^{\mu}_{A A'}$ satisfies the relation
\begin{equation}
\left( \sigma^{\mu}_{A A'}	\sigma^{\nu}_{B B'}+
\sigma^{\nu}_{A A'}	\sigma^{\mu}_{B B'} \right)\varepsilon^{A' B'} = g^{\mu\nu} \varepsilon_{A B},
\end{equation}
where $\varepsilon_{AB}$ and $\varepsilon_{A'B'}$ are the Levi-Civita symbols in two dimensions.

Now, we are going to use this object to write down any tensor with Greek indices, $V_{A A'} = \sigma^{\mu}_{A A'} V_{\mu}$. This also applies to replacing Greek indices with Latin indices. We have 
\begin{align}
\sigma^{\mu}_{A A'} &= (e^{-1})^{\mu}_{a} \sigma^{a}_{A A'},     & \sigma^a&=\frac{1}{\sqrt{2}} (\mathbb{I}, \sigma^i),
\end{align}
where $\sigma^{i}$ are the Pauli matrices and $(e^{-1})^{\mu}_{a}$ is the inverse vierbein defined by
\begin{equation}
g^{\mu\nu} =(e^{-1})^{\mu}_{a} (e^{-1})^{\nu}_{a} \eta^{a b}.
\end{equation}
Here we are using the convention $\eta^{a b} =$ diag$(2,2)$. With $\sigma^{\mu}_{A A'}$ we obtain $\partial_{a}=\sigma_{\mu}^{A A'}\partial_{A A'}$, this will be useful below in eq.~\eqref{MaxwellEc}.

Following with the contraction of eq.~\eqref{WDC2} with four spinors $\xi^A$, we have 
\begin{equation}
	\Psi_{ABCD}\xi^A\xi^B\xi^C\xi^D =\frac{\Phi^{(1)}_{A B} \Phi^{(2)}_{CD}}{\Phi} 	\xi^A\xi^B\xi^C\xi^D. 
\label{DoubleCopyPoli}
\end{equation}
Let $\xi^A = (1 , Z)$, then $\Psi_{ABCD}\xi^A\xi^B\xi^C\xi^D$ is a degree 4 polynomial in variable $Z$. Therefore, it can be written as
\begin{equation}
\label{WeylPoli}
	\Psi_{ABCD}\xi^A\xi^B\xi^C\xi^D = \alpha (x^{a}) \prod^4_{i=1}P_i(Z),
\end{equation}
where  $P_i(Z)$ is a monomial and $\alpha (x^a)$ is a function of the coordinates $x^a= \{x,y,u,v \}$. On the other hand, $\Phi_{A B}\xi^A\xi^B $ is also a polynomial in $Z$, but up to degree 2. Thus, we have
\begin{equation}
\Phi^{(i)}_{A B}\xi^A\xi^B = h_i (x^a) \prod^2_{k=1}P_k(Z),
\label{Polinomio2Maxwell}
\end{equation}
where $i\in \{1,2\}$, and $h_1(x^a)$, $h_2(x^a)$ are also functions of the coordinates. Using eq.~\eqref{Polinomio2Maxwell} we get
\begin{equation}
\frac{\Phi^{(1)}_{A B} \Phi^{(2)}_{CD}}{\Phi} 	\xi^A\xi^B\xi^C\xi^D = \frac{h_1(x^a)h_2(x^a)}{\Phi}\prod^4_{i=1}P_i(Z),
\label{MaxwellPoli}
\end{equation}
then, from eqs. (\ref{WeylPoli}) and (\ref{MaxwellPoli}), we have the following implication
\begin{equation}
\label{relationship}
\frac{h_1(x^a) h_2(x^a) }{\Phi}	 =  \alpha (x^a).
\end{equation}
On the other hand,  any Maxwell spinor can be written as the symmetrized product of two spinors $\textbf{r}_A$ and $\textbf{s}_A$
\begin{equation}
\label{spinorMaxwell}
\Phi^{(i)}_{A B}=h_i(x^a) \textbf{r}_{(A} \textbf{s}_{B)} = h_i(x^a) \left(
\begin{array}{cc}
 A_{00} & A_{01}\\
 A_{10} & A_{11} \\
\end{array}
\right),
\end{equation}
where $i \in \{1,2 \}$ and 
\begin{equation}
A_{00} = r_0s_0, \ \ \ \ \ A_{01} = {1 \over 2} (r_0s_1 +r_1s_0)=A_{10}, \ \ \ \ \ A_{11}=r_1s_1.
\end{equation}
Then, in terms of $Z$ from $\xi^A =(1,Z)$, we have
\begin{equation}
\label{SpinorGeneral2}
\Phi^{(i)}_{A B}\xi^A\xi^B=h_i(x^a) (r_0 + r_1 Z)(s_0 + s_1 Z).
\end{equation}
Comparing eqs.~\eqref{Polinomio2Maxwell} and~\eqref{SpinorGeneral2}, we find the coefficients of eq. (\ref{spinorMaxwell}).

Now, in spinorial formalism, source-free Maxwell's equations are given by \cite{Penrose:1986ca}
\begin{equation}
\label{MaxwellEc}
\partial_{A B'} \Phi^{(i) A B}= 0,
\end{equation}
where $\partial_{\mu}=\sigma_{\mu}^{A A'}\partial_{A A'}$ is the partial derivative. In the flat spacetime  $\partial_{A A'} =\left( \frac{\partial}{\partial x^A}, \frac{\partial}{\partial u^A} \right)$, where $\frac{\partial}{\partial u^A}=\left( {\frac{\partial}{\partial u},\frac{\partial}{\partial v}} \right)$. The spinor $\Phi^{(i)AB}$ is the Maxwell (or electromagnetic) spinor, which represents the electromagnetic field strength. 
With the aid of eq.~\eqref{spinorMaxwell},  eq.~\eqref{MaxwellEc} reads
\begin{align}
\label{1Maxwell2}
\begin{pmatrix}
 A_{10} \\
 A_{11}
\end{pmatrix} \partial_y h_i(x^a)+ h_i(x^a) \partial_y
\begin{pmatrix}
 A_{10} \\
 A_{11}
\end{pmatrix} +
\begin{pmatrix}
 A_{00} \\	
 A_{01}
\end{pmatrix} \partial_x h_i(x^a)+ h_i(x^a) \partial_x
\begin{pmatrix}
 A_{00} \\
 A_{01}
\end{pmatrix}
    =0,  \\
\label{2Maxwell2}
\begin{pmatrix}
 A_{10} \\
 A_{11}
\end{pmatrix} \partial_u h_i(x^a)+ h_i(x^a) \partial_u
\begin{pmatrix}
 A_{10} \\
 A_{11}
\end{pmatrix} -
\begin{pmatrix}
 A_{00} \\	
 A_{01}
\end{pmatrix} \partial_v h_i(x^a)- h_i(x^a) \partial_v
\begin{pmatrix}
 A_{00} \\
 A_{01}
\end{pmatrix}
    =0.
\end{align}
Eqs.~\eqref{1Maxwell2} and~\eqref{2Maxwell2} are differential equations for $h_i=h_i(x,y,u,v)$ and they are equal if  $\Phi^A{ }_{B}$'s are equal.

 In the case of Maxwell's equations with sources we have
\begin{equation}
\partial_{A B'} \Phi^{(i) A B}= J^{B}{ }_{B'},
\label{MaxwellSources}
\end{equation}
we use eq.~\eqref{relationship} in order to find $h_i(x^a)$ and it is possible to verify that they satisfy the eq.~\eqref{MaxwellEc}, otherwise, we solve the differential eq.~\eqref{MaxwellSources} to find $h_i(x^a)$.





\subsection{Hyper-Hermitian Weyl double copy: Preliminaries}
In this subsection, we give a brief overview of the family of solutions called elementary states introduced in ref. \cite{Dunajski:1998nj}. Let $x^{AA'}=(x^A,u^A)$ be local null coordinates on a complex four-manifold $\mathcal{M}$ and let $\Theta_A$ be a pair of complex-valued functions (potentials) on $\cal{M}$, we take $x^A=(y,-x)$ and $u^A=(u,v)$. We write down the eqs. (\ref{igualdad}) once again for convenience
$$
\partial^{2}\Theta^{A}=\{\Theta^{B},\partial_{B}\Theta^{A}\} =0.
$$
As we pointed out in subsection (2.1) eq. (\ref{igualdad}) can be reduced to the second heavenly equation in \eqref{Plebanski} if $\partial_{A0'} \Theta^A=0$. The previous condition implies that $\Theta^A = \partial^A_{ \ 0'} \theta$, where $\theta$ is a holomorphic function which satisfies the Pleba\'nski's second heavenly equation \eqref{Plebanski}, (see \cite{Dunajski:1998nj}).

A hyper-Hermitian manifold has the conformal curvature given by a (anti) self-dual Weyl tensor. The description of its (anti) self-dual Weyl curvature spinor $\Psi_{ABCD}$ in terms of $\Theta_A$ is given by \cite{Dunajski:1998nj}
\begin{equation}
\Psi_{ABCD} =\partial_{(A}\partial_B\partial_{C} \Theta_{D)}.	
\label{WeylSpinor}
\end{equation} 

In the present article, we will be interested in a family of solutions to eq. (\ref{igualdad}) of the hyper-Hermitian metrics called {\it elementary states} \cite{Dunajski:1998nj}, given by
\begin{equation}
\Theta_A=\frac{1}{u x + y v} F_A\left( \frac{u}{u x + y v},\frac{v}{u x + y v}\right), 
\label{twofunctionstheta}
\end{equation}
where $F_A$ are two arbitrary functions of $u^B/(u x + y v)$. In terms of functions $F_A$, a non-singular rescaled hyper-Hermitian metric is written as
\begin{multline}
ds^2= d \left( {x_A \over u x + y v}\right) d\left( {u^A \over u x + y v} \right) +\bigg[ F_B + {u^C \over (u x + y v)}{\partial F_B \over \partial \big({u^C \over u x + y v}\big) }  \bigg] 
\bigg[\frac{u_A u^A}{(u x +y v)^2} d\left( {u^B \over u x + y v}\right)\\   - {u^B \over (u x + y v)} d\left( \frac{x_A u^A}{(u x +y v)^2} \right) \bigg]\bigg[ \frac{u_A}{u x +y v} d\left( \frac{u^A}{u x +y v}\right) \bigg].
\end{multline}

Then, the class of solutions of our interest in the present article for the components of the $F_A$ functions will be of the polynomial form 
\begin{equation}
\label{General}
F_A = \left( a_1 \frac{u^{a_2}v^{a_3}}{(ux+yv)^{a_4}}, b_1  \frac{u^{b_2}v^{b_3}}{(ux+yv)^{b_4}} \right),
\end{equation}
where $a_i$ and $b_i$, with $i \in \{1,2,3,4\}$, are constants.
In order for eqs.~\eqref{twofunctionstheta} with~\eqref{General} satisfy $\partial^2 \Theta_C=0$, the parameters of that solution must take certain values. For the equation $\partial^2 \Theta_C=0$ to be satisfied, it is necessary to choose at least one of the three conditions from the set of parameters $\{a_1=0, a_4=a_2+a_3, a_4=-1 \}$ and at least one condition
from the set $\{b_1=0, b_4=b_2+b_3, b_4=-1 \}$. Thus, in total, we will have nine cases. For instance, the selection of $a_1=0$ and $b_1=0$ yields the trivial solution, the other eight cases give nontrivial ones. However, three cases will be repeated due to the symmetry under the interchange of parameters $a \leftrightarrow b$. One of these cases is $\{a_1=0, b_4=-1\}$ and $\{a_4=-1, b_1=0\}$.

These conditions are necessary for $F_A$ to be elementary states. It is important to remark that due to the symmetry between parameters $a_i$ and $b_i$ in $F_A$, many of the choices made are repeated and therefore we carry out the analysis restricting to take just one set of parameters.

With eq. (\ref{General}) the contracted Weyl curvature spinor is given by
\begin{equation}
\Psi_{A B C D}\xi^{A} \xi^{B} \xi^{C} \xi^{D}=\frac{-1}{(u x+y v)^{4}}(v-u Z)^{3}\bigg[g_{a_{1} a_{2} a_{3} a_{4}}(x^a)+g_{b_{1} b_{2} b_{3} b_{4}}(x^a) Z\bigg],
\end{equation}
where
\begin{equation}
g_{a_{1} a_{2} a_{3} a_{4}}(x^a) \equiv a_{1}\left(1+a_4 \right)\left(2+a_{4}\right)\left(3+a_4 \right) u^{a_{2}} v^{a_{3}}(u x+y v)^{-a_{4}},
\end{equation}
where there is no restriction on the set of parameters. Thus, for generic hyper-Hermitian manifolds, we can see that the Weyl curvature has three equal roots  and the other different and consequently we get a type {\bf III} metric of the Petrov classification.

Now, in order to fulfil the following condition
\begin{equation}
\partial_A \Theta^A = 0,
\end{equation}
we set $a_1=0$ and $b_4=b_2+b_3$. Another possibility is to take
\begin{align}
&b_2=a_2+1, & a_3 = b_3+1, \nonumber \\
& b_4=a_4=b_2+b_3, & a_1=-b_1.
\label{ConditionKahler}
\end{align}

On the other hand, starting also from the general case with unrestricted parameters we establish the conditions under which Weyl spinor leads to type {\bf N} solutions. Thus one has  
\begin{align}
\Psi_{A B C D} \xi^{A} \xi^{B} \xi^{C} \xi^{D} & =\frac{-1}{(u x+y v)^{4}}(v-u Z)^{3}\left(g_{a_{1} a_{2} a_{3} a_{4}}+g_{b_{1} b_{2} b_{3} b_{4}} Z\right) \nonumber \\
& =\frac{-1}{(u x+y v)^{4}}(v-u Z)^{3} \frac{g_{a_{1} a_{2} a_{3} a_{4}}}{v}\left(v+\frac{v g_{b_{1} b_{2} b_{3} b_{4}}}{g_{a_{1} a_{2} a_{3} a_{4}}} Z\right),
\label{weyl}
\end{align}
with
\begin{equation}
-u=\frac{v g_{b_{1} b_{2} b_{3} b_{4}}}{g_{a_{1} a_{2} a_{3} a_{4}}}. 
\label{Ncondition}
\end{equation}
Recall that a solution of the vacuum Einstein equations is type {\bf N} metric if it has four principal null directions equal, and from eqs. (\ref{weyl}) and (\ref{Ncondition}). It happens if the following relations are fulfilled
\begin{align}
&b_2=a_2+1, & a_3 = b_3+1,\nonumber \\
 & b_4=a_4=b_2+b_3, & a_1=-b_1.
\end{align}
Those conditions are precisely the same that we found in eq.~\eqref{ConditionKahler}. Thus, we  have that the result {\it elementary states have to Petrov  type} \textbf{N} {\it metric } is equivalent to the condition {\it elementary states have to be hyper-K\"ahler}.  We summarize these results in Table \ref{tab:Parameters} 
in which we omit the parameter values that give us the same metric, for example, $a_1=0$ and $b_4=b_2+b_3$,  give the same metric as $a_4=-1$ and $b_4=b_2+b_3$.
\begin{table}[]
\centering
\resizebox{\textwidth}{!}{%
\begin{tabular}{@{}cll@{}}
\toprule
\multicolumn{1}{l}{}           & \multicolumn{1}{c}{Type {\bf III}}                                                                                                                 & \multicolumn{1}{c}{Type {\bf N}}                             \\ \midrule
 Hyper-Hermitian                & \begin{tabular}[c]{@{}l@{}}$a_1=0$, $b_4=b_2+b_3$.\\  $a_4=a_2+a_3$, $b_4=b_2+b_3.$\end{tabular} &                                                        \\
Hyper-K\"ahler &                                                                                                                                             & $a_4=a_2+a_3$, $b_4=b_2+b_3$, $b_2=a_2+1$, $a_3=b_3+1$, $a_1=-b_1$. \\ \bottomrule
\end{tabular}
}
\caption{In the table is depicted the parameters that allow to build elementary states and how they are ordered according to how they give rise to hyper-Hermitian or hyper-K\"ahler solutions. For the hyper-Hermitian case they generically lead to type {\bf III} solutions and the hyper-K\"ahler give type {\bf N} solutions. For the hyper-K\"ahler case we restricted the allowed of the parameters from eqs.~\eqref{twofunctionstheta} and \eqref{General}, and then impose separately the implementation of conditions $\partial^2 \Theta_A=0$ and $\partial_A \Theta^A$=0.}
\label{tab:Parameters}
\end{table}

\subsection{Hyper-Hermitian Weyl double copy: Generic case}
In the present subsection, we provide a series of examples of the hyper-K\"ahlerian and the hyper-Hermitian double copy correspondence. We start from the mentioned case of elementary states encoded in the family of solutions given by eqs.~\eqref{twofunctionstheta} and \eqref{General}
\begin{equation}
\Theta_A = \left( a_1 \frac{u^{a_2}v^{a_3}}{(ux+yv)^{a_4+1}}, b_1  \frac{u^{b_2}v^{b_3}}{(ux+yv)^{b_4+1}} \right).
\label{General0}
\end{equation}

The contracted spinorial Weyl curvature can be written by using eq. (\ref{DoubleCopyPoli}) as follows
\begin{align}
 \Psi_{ABCD}\xi^A\xi^B\xi^C\xi^D  &= \frac{-1}{(u x+y v)^{4}}(v-u Z)^{3}\big[g_{a_{1} a_{2} a_{3} a_{4}}(x^a)+g_{b_{1} b_{2} b_{3} b_{4}}(x^a) Z\big]  \nonumber \\ 
&=  \frac{\Phi^{(1)}_{A B} \Phi^{(2)}_{CD}}{\Phi} 	\xi^A\xi^B\xi^C\xi^D.
\label{WeylDCC}
\end{align}
Thus, with the aid of eq. (\ref{WeylPoli}), we can read off factor $\alpha$ and the polynomial factors
\begin{equation}
\alpha(x^a) = -(u x+y v)^{-4},
\end{equation}
\begin{equation}
P_1(x^a)=P_2(x^a)=P_3(x^a)= v-u Z,
\end{equation}
and
\begin{equation}
P_4(x^a)= g_{a_{1} a_{2} a_{3} a_{4}}(x^a)+g_{b_{1} b_{2} b_{3} b_{4}}(x^a) Z.
\end{equation}
Moreover, we know that $\Phi^{(1/2)}_{A B}\xi^A\xi^B$ is a polynomial in the variable $Z$ up to degree 2, but in principle, we do not know how to find the factor $\alpha (x^a)$ between the two contracted Maxwell spinors. Therefore, the field strength spinors are
\begin{align}
\Phi^{(1)}_{AB}\xi^A\xi^B &= (v-u Z)^2 h_1(x^a), \\
\Phi^{(2)}_{AB}\xi^A\xi^B &= (v-u Z)\big[g_{a_{1} a_{2} a_{3} a_{4}}(x^a)+g_{b_{1} b_{2} b_{3} b_{4}}(x^a) Z\big] h_2(x^a),
\end{align}
where $h_i(x^a)$ represent our lack of knowledge about the factor $\alpha(x^a)$. Now, from the comparison between  eqs.~\eqref{spinorMaxwell} and~\eqref{SpinorGeneral2}, we can find the spinor entries
\begin{equation}
\Phi_{AB}^{(1)}=
\left(
\begin{array}{cc}
 v^2 & -u v \\
 -u v & u^2 \\
\end{array}
\right) h_1(x^a).
\end{equation}
Since the polynomial $(u-v Z)^2$ is of the same form as for the Maxwell spinor in Sparling-Tod double copy in eq. (\ref{ReferenciaFinal}) (see below), we establish that
\begin{equation}
h_1(x^a) = \sqrt{\frac{\Phi}{(u x + y v)^5}},
\end{equation}
where
\begin{equation}
\Phi=A \frac{1}{(u x + yv)}+B_r \frac{ \left(u x+y v\right)^r}{v^2 u^{r-1}} +C_r \frac{ \left(u x+y v\right)^r}{u^2 v^{r-1}}
\label{General3}
\end{equation}
for all $r \in \mathbb R$ and where $A$, $B_r$ and $C_r$ are constants as we mentioned previously. Finally, the explicit form of the Maxwell spinor is written as
\begin{equation}
\Phi_{AB}^{(1)}=
\left(
\begin{array}{cc}
 v^2 & -u v \\
 -u v & u^2 \\
\end{array}
\right) \sqrt{\frac{\Phi}{(u x + y v)^5}}.
\label{General2}
\end{equation}
Eq.~\eqref{General2} satisfies eq.~\eqref{MaxwellEc}. Analogously, the field strength spinor $\Phi_{AB}^{(2)}$ is given by
\begin{equation}
\Phi_{AB}^{(2)}=
\left(
\begin{array}{cc}
 v g_{a_1 a_2 a_3 a_4}(x^a)& \frac{1}{2}(v g_{b_1 b_2 b_3 b_4}(x^a)-u g_{a_1 a_2 a_3 a_4}(x^a) ) \\
\frac{1}{2}(v g_{b_1 b_2 b_3 b_4}(x^a)-u g_{a_1 a_2 a_3 a_4}(x^a) ) & -u g_{b_1 b_2 b_3 b_4}(x^a) \\
\end{array}
\right) h_2(x^a),
\label{SpinorMaxwell2}
\end{equation}
where
\begin{equation}
h_2(x^a)=-\sqrt{\frac{\Phi}{(u x + y v)^{3}}}.
\end{equation}
One can check that eq. (\ref{SpinorMaxwell2}) satisfies eq. (\ref{MaxwellSources}), where the current $J^{AA'}$ is not conserved in the following form
\begin{equation}
\partial_{A A'} J^{A A'} =
\frac{u p_{a_1 a_2 a_3 a_4}(x^a) q_{a_1 a_2 a_3 a_4}(x^a)  + v  p_{b_1 b_2 b_3 b_4}(x^a)  r_{b_1 b_2 b_3 b_4}(x^a) }{2 u^2 (u x+ y v)^{-3+a_4+b_4}},
\label{nonconserva}
\end{equation}
where
\begin{equation}
 p_{a_1 a_2 a_3 a_4}(x^a) \equiv a_1 (a_4+1)(a_4+2)(a_4+3)u^{a_2} v^{a_3}(u x +y v)^{a_4},
\end{equation}
\begin{align}
&q_{a_1 a_2 a_3 a_4}(x^a)\equiv u^2 [(2 + a_4) (3 + a_4) u^2 +  2 (1 + a_3) (2 + a_4) u x \nonumber \\
&- (-2 + a_2 - a_4) (1 + a_2 + 2 a_3 - a_4) x^2 ] - 2 u v [(2 + a_4) (2 - a_3 + a_4) u \nonumber \\
&+ (2 - 2 a_3 + a_2 (-1 + a_2 + 2 a_3) + 3 a_4 - (2 a_2 + a_3) a_4 + a_4^2) x] y 
\nonumber \\&- a_2 (-1 + a_2 + 2 a_3 - 2 a_4) v^2 y^2\end{align}
and
\begin{align}
&r_{b_1 b_2 b_3 b_4}(x^a)\equiv u^2 [(2 + b_4) (3 + b_4) u^2 +   2 (2 + b_4) (2 - b_2 + b_4) u x 
\nonumber \\&+ (2 - b_2 + b_4) (3 - b_2 + b_4) x^2] -   2 u v [(1 + b_2) (2 + b_4) u  \nonumber \\
&+ b_2 (3 - b_2 + b_4) x] y + (-1 + b_2) b_2 v^2 y^2.
\end{align}
Thus, generically in the hyper-Hermitian case, besides the existence of two different Maxwell spinors, one of them will be associated with a source current and the other will be sourceless. The process in which the Weyl double copy includes a source in the gauge sector has been studied in refs. \cite{Easson:2021asd, Easson:2022zoh}. We reproduce its result, where one Maxwell spinor is sourceless, and at least one Maxwell spinor has a source. This procedure will be the same for all the following examples that will be described in the next section. This is, therefore, one of our main results: for the Weyl double copy in Hermitian manifolds and elementary states in general, it holds that
\begin{equation}
 \Psi_{ABCD} =  \frac{\Phi^{(1)}_{A B} \Phi^{(2)}_{CD}}{\Phi},
\end{equation}
where $\Psi_{ABCD}$ is given by eq.~\eqref{WeylDCC}, $\Phi$ corresponds to eq.~\eqref{General3} and $\Phi_{AB}^{(1)}$ and $\Phi_{AB}^{(2)}$ are given by eqs.~\eqref{General2},~\eqref{SpinorMaxwell2} respectively with $\{ a_1=0$, $b_4=b_2+b_3 \}$ given one metric or $\{a_4=a_2+a_3$, $b_4=b_2+b_3\}$ given other metric. We called these results \textit{generic hyper-hermitian}  double copy. We note that, since the equation $\partial^2 \Theta_A = 0$ is linear, we can take linear combinations of the potential in eq.~\eqref{General0} without any issue, which highlights the generality of the result.

\section{Examples}
\label{examples}
In this Section, we will give several examples regarding our results of the previous section. We will discuss the explicit Weyl double copies for five examples. Two of them will lead to the hyper-K\"ahler double copy case and the rest will be genuine hyper-Hermitian.
\subsection{Hyper-K\"ahler double copy: Generic case}
\label{Example1}
In what follows, we will consider the solutions $\Theta_A$ of the form given by eq.~\eqref{General0} with $a_4=a_2+a_3$, $b_4=b_2+b_3$, $b_2=a_2+1$, $a_3=b_3+1$, $a_1=-b_1$. Thus we start from the following pair of potentials 
\begin{equation}
\label{Ex15}
\Theta_D= \left( \frac{a_1 u^{b_2-1} v^{b_3+1}}{(u x+y v)^{b_2+b_3+1}},\frac{-a_1 u^{b_2} v^{b_3}}{(u x+y v)^{b_2+b_3+1}} \right). 
\end{equation}
 The metric obtained from these potentials is given by
\begin{equation}
\label{metricGeneral2}
d{s}^2 = d{u}d{x}+d{y}d{v}- \frac{a_1(b_2+b_3+1) u^{b_2-1} v^{b_3}}{(u x+y v)^{b_2+b_3+2}}  (u dv- v du)^2.
\end{equation}
In this case, the contracted Weyl spinor reads 
\begin{equation}
\Psi_{ABCD}\xi^A\xi^B\xi^C\xi^D = c u^{b_2-1} v^{b_3} (u x+y v)^{-b_2-b_3-4} (v-u Z)^4,
\end{equation}
with 
$$
c\equiv -a_1 (1+b_2+b_3)(2+b_2+b_3)(3+b_2+b_3).
$$
Moreover, we identify the polynomial factors which are given by
\begin{equation}
P_1(Z)=P_2(Z)=P_3(Z)=P_4(Z)= v-uZ,
\end{equation}
and
\begin{equation}
\alpha (x^a)=c u^{b_2-1} v^{b_3} (u x+y v)^{-b_2-b_3-4}.
\end{equation}
We have a type {\bf N} metric, then the field strength spinor has the form
\begin{equation}
\Phi^{(1)}_{AB}=\Phi^{(2)}_{AB}=\Phi_{AB}=\left(
\begin{array}{cc}
 v^2 & -u v \\
 -u v & u^2 \\
\end{array}
\right) h(x^a),
\end{equation}
thus, in this case, both Maxwell spinors are equal. Now, imposing eq.~\eqref{relationship} the function $h(x^a)$ can be written in the form
\begin{equation}
\label{eq15f}
h(x^a)=c_1 \sqrt{u^{b_2-1} v^{b_3} (u x+y v)^{ -b_2-b_3-4} \Phi},
\end{equation}
where $\Phi$ is given by eq.~\eqref{ZeroCopy} and $c_1 \in \mathbb{R}$ is absorbed in the constants of $\Phi$. Writing explicitly the only Maxwell spinor we have
\begin{equation}
\Phi_{AB}=\left(
\begin{array}{cc}
 v^2 & -u v \\
 -u v & v^2 \\
\end{array}
\right)\sqrt{u^{b_2-1} v^{b_3} (u x+y v)^{ -b_2-b_3-4} \Phi}.
\label{MaxwellField}
\end{equation}
Electric field obtained from eq. (\ref{MaxwellField}) is plotted  in Figure 1, where we select the parameters of the form:  $c_1=A=r=b_2=1$ and $t=B_r=C_r=b_3=z=0$.

It is easy to see also that Sparling-Tod solution \cite{Sparling:1981nk} can be recovered from the following values of the parameters with $b_2=1$ and $b_3=0$, thus we get
\begin{equation}
\Psi_{ABCD}\xi^A\xi^B\xi^C\xi^D = c (u x+y v)^{-5} (v-u Z)^4. 
\end{equation}
Moreover, it is possible to see that eq.~\eqref{eq15f} becomes 
\begin{equation}
h(x^a)= \sqrt{\frac{c_1 \Phi} {(u x+y v)^{5 }}},
\end{equation}
which is the same as the one from eq.~\eqref{fSparlingTod} (see below). The scalar $\Phi$ is given by eq.~\eqref{ZeroCopy}.

\begin{figure}[h]

\hfill
\begin{minipage}[t]{.90\textwidth}
\begin{center}
\includegraphics[scale=0.65]{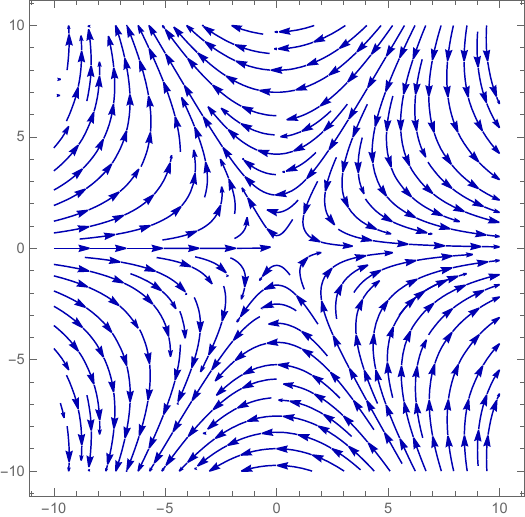}
\end{center}
\end{minipage}
\hfill
\caption{In this figure the electric field obtained from the Maxwell field strength in eq. (\ref{MaxwellField}) of subsection \ref{Example1}. Some parameters are selected just with $c_1=A=r=b_2=1$ and $t=B_r=C_r=b_3=z=0$.}
\hfill
\label{figureEx2}
\end{figure}


To visualize the solution we plot the electric field obtained from eq. (\ref{MaxwellField}) in Figure \ref{figureEx2}, in terms of Minkowskian coordinates  $(x^0,x^1,x^2,x^3)$ which are defined by
\begin{align} 
u &= \frac{i}{\sqrt{2}} ( x^0 - x^3),   \ \ \ \ \ \ \ 
x = \frac{i}{\sqrt{2}} ( x^0 + x^3) ,  \nonumber \\
y &= -\frac{1}{\sqrt{2}} (i x^1 - x^2),  \ \ \ \ \ \ \ v = \frac{1}{\sqrt{2}} (i x^1 + x^2).  
\label{ccordinatesT}
\end{align}
We observe that the solution in the gauge sector is self-dual.

\subsection{Sparling-Tod Double Copy}
\label{sec_SparlingTodDC}
\begin{figure}[h]
\centering 
\begin{subfigure}{.45\textwidth}
\centering
\includegraphics[scale=0.65]{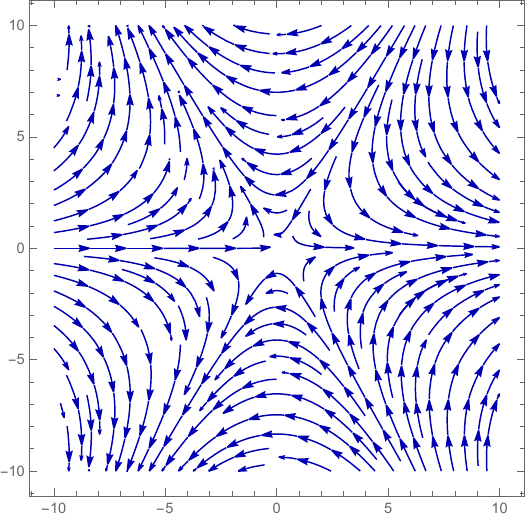}
\caption{Here, the selected parameters are: $A=1$, $B_r=0$, $C_r=0$, $z=0$.}
\end{subfigure} 
\hfill
\begin{subfigure}{.45\textwidth}
\centering
\includegraphics[scale=0.65]{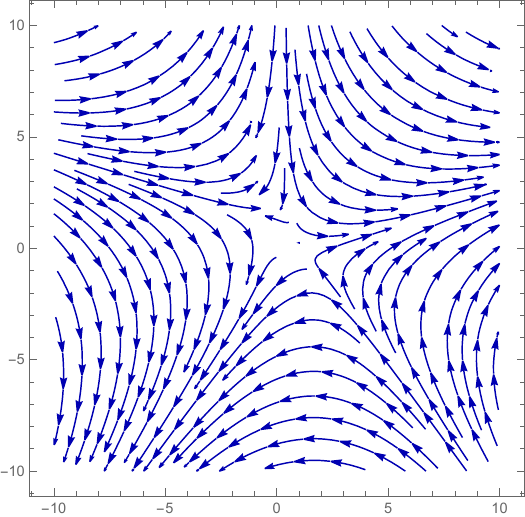}
\caption{The parameters used are: $A=1$, $B_r=1$, $C_r=1$, $r=2$, $z=1$ and $t=0$.}
\end{subfigure}
\caption{In these figures, we plot the Maxwell spinor in eq.~\eqref{spinorST} corresponding to  Sparling-Tod's electric field with different values of parameters in $\Phi$.    }
\hfill
\label{figureST}
\end{figure}

 In ref.  \cite{Dunajski:1998nj} it was discussed the case of Sparling-Tod spacetime
\cite{Sparling:1981nk}, which is also a heavenly or self-dual space ($\mathfrak{H}$-space). For the purpose of regaining the Sparling-Tod self-dual metric in the $\Theta_A$-formalism, one must take 
\begin{equation}
\label{STeq}
\Theta_D= {1 \over (u x + y v)^2} (-v,u).
\end{equation} 
The Sparling-Tod metric can be obtained from  eq.~\eqref{Ex15} for the following specific values of parameters: $b_3=0$, $b_2=1$, $a_1=-1$. Thus, the metric is expressed as
\begin{equation}
d{s}^2= d{u}d{x} + d{y}d{v} + 2 \frac{( u d{v}- v d{u})^2}{(u x+y v)^3} .
\end{equation}
In this case knowing that this solution is type {\bf N} metric, the contracted Weyl curvature reads
\begin{equation}
\Psi_{ABCD}\xi^A\xi^B\xi^C\xi^D = \frac{24 (v-u Z)^4}{(u x+y v)^5}.
\label{STsol}
\end{equation}
From eq.~(\ref{WeylPoli}), one can be read off the polynomials
\begin{equation}
P_1(Z)=P_2(Z)=P_3(Z)=P_4(Z)= v-uZ,
\end{equation}
where we make the identification 
\begin{equation}
\alpha (x^a) = \frac{24}{(ux+yv)^5}.
\end{equation}
Then, we find the Maxwell spinors in the following form 
\begin{align}
\Phi^{(1)}_{AB}\xi^A\xi^B &= (v-u Z)^2 h_1(x^a), \\
\Phi^{(2)}_{AB}\xi^A\xi^B &= (v-u Z)^2 h_2(x^a).
\label{ReferenciaFinal}
\end{align}
Explicitly, we find that there is only one Maxwell spinor, and it is given by
\begin{equation}
\Phi_{AB}=
\left(
\begin{array}{cc}
 v^2 & -u v \\
 -u v & u^2 \\
\end{array}
\right) h(x^a),
\end{equation}
where, with the aid of eq.~\eqref{relationship}, we obtain 
\begin{equation}
\label{fSparlingTod}
h(x^a)=h(u,x,y,v)= \sqrt{ \frac{24 \Phi}{(ux+yv)^5}}.
\end{equation}
Here $\Phi$ is given by eq.~\eqref{ZeroCopy} and it reproduces Sparling-Tod metric when $\Phi=(u x+y v)^{-1}$. With these ingredients, it is possible to find the explicit form of Maxwell spinor as follows:
\begin{equation}
\label{spinorST}
\Phi_{AB}=
\left(
\begin{array}{cc}
 v^2 & -u v \\
 -u v & u^2 \\
\end{array}
\right) \sqrt{ \frac{24 \Phi}{(ux+yv)^5}},
\end{equation}
where
\begin{equation}
\Phi=A \frac{1}{(u x + yv)}+B_r \frac{ \left(u x+y v\right)^r}{v^2 u^{r-1}} +C_r \frac{ \left(u x+y v\right)^r}{u^2 v^{r-1}},
\end{equation}
for all $r \in \mathbb R$ and where $A$, $B_r$ and $C_r$ are constants as we already mentioned. 
These results with different families of solutions and only one metric were obtained similarly in ref. \cite{Gonzo:2021drq}. An example of a type \textbf{D} metric is the Eguchi-Hanson metric, as described in ref. \cite{Dunajski:1998nj}, which is obtained with an Euclidean slice. We plot in Figure \ref{figureST} eq.~\eqref{spinorST} with different values of the parameters.

So far, the results we have obtained are  type {\bf N} metrics. This is expected since the two metrics are hyper-Kähler. In the following examples, we see metrics that are not hyper-Kähler and with leading null directions that have a multiplicity different from 4, all these results are summarized in the Tables \ref{tab:ResumeResults} and \ref{tab:ResumeResults2}. 

\subsection{Potentials with $\Theta_0=0$}
\label{Example2}
Now we consider solutions $\Theta_A$ of the form eq. (\ref{General0}) with $a_1=0$ and $b_4=b_2+b_3$, thus the potential is
\begin{equation}
\Theta_A = \left(0,\frac{b_1 u^{b_2} v^{b_3}}{(u x+v y)^{b_2+b_3+1} } \right),
\label{halftSpinor2}
\end{equation}
therefore we have that the manifold $\cal{M}$ obtained from eq.~\eqref{halftSpinor2} is hyper-Hermitian. The corresponding metric obtained from this potential is given by
\begin{equation}
ds^2= du dx + dy dv + 
 b_1 (1 + b_2 + b_3) dv u^{b_2} v^{b_3} (dv u - du v) (u x + v y)^{-2 - b_2 - b_3},
\end{equation}
if we consider eq. (\ref{General0}) with the parameters $a_4=-1$ and $b_4=b_2+b_3$, we obtain the same metric. So, we only concentrate on these parameters. The same happens for $b_1=0$, $a_4=a_2+a_3$ and $b_4=-1$, $a_4=a_2+a_3$, which being symmetric parameters give us similar solutions. Now we compute the contracted Weyl curvature, which reads
\begin{multline}
\Psi_{ABCD} \xi^A\xi^B\xi^C\xi^D \\ = b_1 (1 + b_2 + b_3) (2 + b_2 + b_3) (3 + b_2 + b_3) u^{b_2} v^{b_3} (u x + 
   v y)^{-4 - b_2 - b_3} Z (-v + u Z)^3.
\end{multline}
We have a type {\bf III} metric, then the double copy in eq.  (\ref{WDC2}), is satisfied when Maxwell spinors are given by
\begin{equation}
\Phi^{(1)}_{AB}=
\left(
\begin{array}{cc}
 v^2 & -u v \\
 -u v & u^2 \\
\end{array}
\right) \sqrt{ \frac{ \Phi}{(ux+yv)^5}},
\end{equation}
and
\begin{multline}
\Phi_{AB}^{(2)}=
\begin{pmatrix}
0 & -\frac{1}{2}  v \\
-\frac{1}{2}  v & u
\end{pmatrix}  \\
\times b_1 (1 + b_2 + b_3) (2 + b_2 + b_3) (3 + b_2 + b_3) u^{b_2} v^{b_3} (u x + v y)^{-2 - b_2 - b_3},
\end{multline}
where we take $\Phi$ with $B_r=C_r=0$ and it satisfies Maxwell's equations with sources 
\begin{equation}
\partial_{A B'} \Phi^{(2) A}{}_{B} = J_{B B'}.
\end{equation}
It is observed that $J_{AA'}$ is not conserved, i.e.
\begin{align}
\partial_{A A'} J^{A A'} &= \frac{1}{4} b_1 (b_2+b_3+1) (b_2+b_3+2) (b_2+b_3+3) u^{b_2-2} v^{b_3+1}  \nonumber\\ & \times(u x+v y)^{-b_2-b_3-4} \left(u^2 \left(u^2 (b_2+b_3+2) (b_2+b_3+3)+ \right.\right. \nonumber\\
&\left. \left. 2 (b_3+2) u x (b_2+b_3+2)+(b_3+2) (b_3+3) x^2\right) \right. \nonumber \\ 
&\left. -2 u v y ((b_2+1) u (b_2+b_3+2)+b_2 (b_3+3) x)+(b_2-1) b_2 v^2 y^2\right).
\end{align}
Notice that this result is consistent with the formula for the generic case in eq. (\ref{nonconserva}) by taking the appropriate values of the parameters.

\begin{figure}[h]
\hfill
\begin{minipage}[t]{.90\textwidth}
\begin{center}
\includegraphics[scale=0.55]{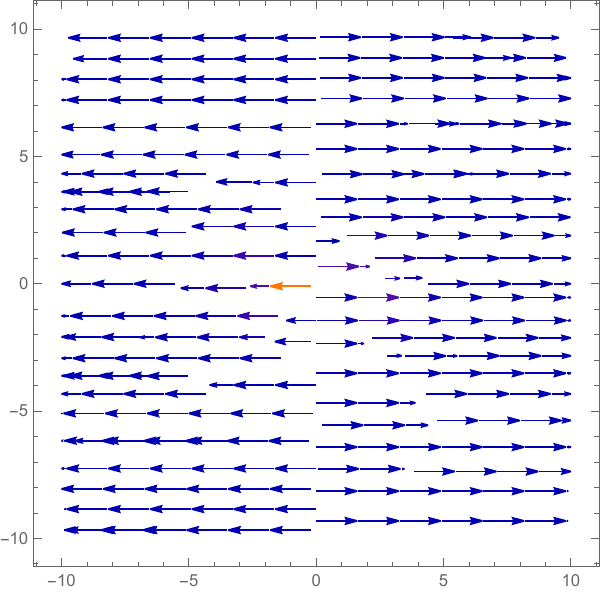}
\end{center}
\end{minipage}
\caption{Electric field of subsection (\ref{Example2}) with $t=0$ and $z=1$. We plotted eq.~\eqref{EqPlotted} depending only on the coordinates of eq.  (\ref{ccordinatesT}). The magnetic field is proportional to the electric field, and therefore it is not plotted.} 
\hfill
\label{figureEx1}
\end{figure}

As a special case, we consider $b_2=1$ and $b_3=0$ to get
\begin{equation}
\Theta_A = \left( 0 , \frac{u}{(u x +y v)^{2}} \right),
\end{equation}
and
\begin{equation}
\Psi_{ABCD} \xi^A\xi^B\xi^C\xi^D = \frac{24 u Z (u Z-v)^3}{(u x+v y)^5},
\end{equation}
then
\begin{equation}
\Phi_{AB}^{(2)}=
-\begin{pmatrix}
0 & u v \\
u v & -2 u^2
\end{pmatrix} \frac{1}{(u x+y v)^3},
\label{EqPlotted}
\end{equation}
and
\begin{equation}
\partial_{A A'} J^{A A'} =  v (u (2 u^2 + 2 u x + x^2) - v (2 u + x) y) ( u x +  v y)^{-5}.
\end{equation}
This solution is hyper-Hermitian and non-hyper-K\"ahler. Some plots of the electric and magnetic fields associated with $\Phi ^{(2)}_{AB}$ are shown in Figure \ref{figureEx1}. The electric and magnetic fields were obtained from eq. (\ref{EqPlotted}) using the flat coordinates  in eq. (\ref{ccordinatesT}).



\begin{table}[]
\centering
\resizebox{12cm}{!}{
\begin{tabular}{ccccccccc}
\hline
\multirow{2}{*}{Metric} & \multicolumn{8}{c}{Principal Null Directions}                                                                                                                                               \\
                        & $\tau_A$                              & m & $\chi_A$                                        & m & $\kappa_A$                              & m & $\omega_A$                              & m \\ \hline
Generic Hyper-K\"ahler            & $\begin{pmatrix} v & -u \end{pmatrix}$ & 4 &  &  &                                        &   &                                        &  
 \\
Sparling-Tod            & $\begin{pmatrix} v & -u \end{pmatrix}$ & 4 &                                                  &   &                                        &   &                                        &   \\
Generic Hyper-Hermitian          & $\begin{pmatrix} v & -u \end{pmatrix}$ & 3 & $ \begin{pmatrix} g_{a_1 a_2 a_3 a_4} (x^a) & g_{b_1 b_2 b_3 b_4} (x^a) \end{pmatrix} $  & 1 &                                        &   &                                        &   \\ 
Potential with $\Theta_0=0$            & $\begin{pmatrix} v & -u \end{pmatrix}$ & 3 & $ \begin{pmatrix} 0 & 1 \end{pmatrix} $  & 1 &                                        &   &                                        &   \\ 
Constant Solution              & $\begin{pmatrix} -v & u \end{pmatrix}$ & 3 & $\begin{pmatrix} 1 & 1 \end{pmatrix} $           & 1 &                                        &   &                                        &   \\ 
Non-elementary Function               & $\begin{pmatrix} 1 & C \end{pmatrix}$  & 1 & $\begin{pmatrix} 1 & -C \end{pmatrix} $          & 1 & $\begin{pmatrix} 0 & 1 \end{pmatrix} $ & 1 & $\begin{pmatrix} 1 & 0 \end{pmatrix} $ & 1 
 \\  \hline
\end{tabular} }
\caption{Principal null directions of the examples in double copy. The multiplicity $m$ is next to the null directions. Non-elementary Function is type {\bf I} metric and Sparling-Tod is type {\bf N} metric.}
\label{tab:ResumeResults}
\end{table}

\subsection{Constant Solution}
\label{Example3}

\begin{figure}[h]
\centering 
\begin{subfigure}{.45\textwidth}
\centering
\includegraphics[scale=0.55]{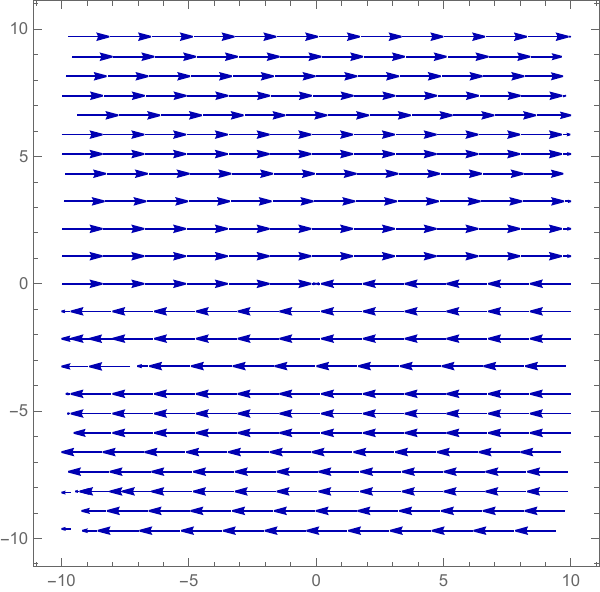}
\caption{Electric field.}
\end{subfigure} 
\hfill
\begin{subfigure}{.45\textwidth}
\centering
\includegraphics[scale=0.494]{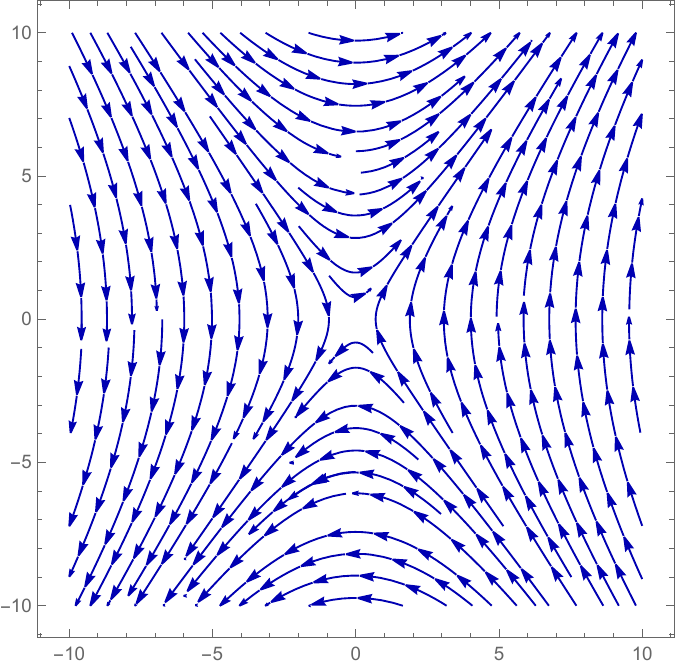}
\caption{Magnetic field.}
\end{subfigure}
\caption{Electric and magnetic field of subsection (\ref{Example3}) corresponding to constant solution with $t=z=0$. To obtain electromagnetic fields we plotted eq. (\ref{EqPlotted2}) in local coordinates of eq. (\ref{ccordinatesT}). }
\hfill
\label{figureEx3}
\end{figure}

In this subsection, we repeat the same procedure as in the previous examples starting again from eq. (\ref{General0}), and taking $a_4=a_3+a_2$, $b_4=b_2+b_3$, $a_1=b_1=1$ and $a_2=a_3=b_2=b_3=0$. The potential in this case is given by
\begin{equation}
\Theta_A = \frac{1}{u x +y v} \left( 1 , 1 \right).
\end{equation}
The hyper-Hermitian metric associated with this potential is
\begin{equation}
d{s}^2 = d{u}d{x}+d{y}d{v}+ \frac{1}{(u x+y v)^2} \bigg[u d{v}^2+  (u -v) d{u}d{v}
- v  d{u}^2\bigg] ,
\end{equation}
which is a type {\bf III} metric. Results where solutions to the Weyl double copy are of type {\bf N }or type {\bf D} have been found in \cite{Mao:2023yle}.

The Weyl spinor curvature of this metric gives rise to
\begin{equation}
\Psi_{ABCD} \xi^{A}\xi^{B}\xi^{C}\xi^{D} = \frac{ 1}{(u x+y v)^4}(Z+1)(u Z-v)^3 .
\end{equation}
In this case, the Maxwell spinors are given by
\begin{equation}
\Phi_{AB}^{(1)}=\left(
\begin{array}{cc}
 v^2 & -u v \\
 -u v & u^2 \\
\end{array}
\right)  \frac{1}{(u x+y v)^3},
\end{equation}
and
\begin{equation}
\Phi_{AB}^{(2)}=
\left(
\begin{array}{cc}
 -v & \frac{1}{2}(u-v) \\
 \frac{1}{2}(u-v) & u \\
\end{array}
\right) \frac{1}{(u x+y v)^2}.
\label{EqPlotted2}
\end{equation}
In Figure \ref{figureEx3} we plot the static electric and magnetic fields in eq. (\ref{EqPlotted2}). They correspond to the constant solution with $t=z=0$.

It is easy to check that $\Phi_{AB}^{(1)}$ satisfies vacuum  Maxwell's equations and $\Phi_{AB}^{(2)}$ satisfies Maxwell's equations with sources $J_{A A'}$ \cite{Penrose:1986ca}
\begin{equation}
J_{A A'}=\left(
\begin{array}{cc}
 2v (u +v) & u x+2 v x- v y \\
 -2 u(u +v) &  -(u x-2 v x- v y) \\
\end{array} 
\right) \frac{3}{(u x +y v)^3},
\end{equation}
where in the present case $J_{AA'}$ is not conserved
\begin{equation}
\partial_{A A'} J^{A A'} =\frac{3 u^3+3 u^2 v+2 u^2 x-2 v y (2 u+v+x)+4 u v x+u x^2+3 v x^2}{(u x+v y)^4}.
\end{equation}
This result can be obtained also from eq. (\ref{nonconserva}) taking the appropriate  choice of the parameters.

We use here a source in the gauge sector considering that Maxwell spinor will not be self-dual.  The electric field is constant and is mapped by the double copy because the function $F_A$ is constant. The scalar field $\Phi$ is given by
\begin{equation}
\Phi = { 1 \over  ( u x + y v)},
\end{equation}
one can see that $\Phi$ can be obtained from eq.~\eqref{ZeroCopy} with $B_r=C_r=0$ and $A=1$. With this we finish our study of the elementary states $F_A$, in the next section, we will study a case that is not an elementary state.

\begin{table}[]
\centering
\begin{tabular}{@{}ccc@{}}
\toprule
Metric      & Type & Classification      \\ \midrule
Generic Hyper-K\"ahler & {\bf N}  & Hyper-K\"{a}hler     \\
Sparlig-Tod & {\bf N}    & Hyper-Kähler \\
Generic Hyper-Hermitian  & {\bf III} & Hyper-Hermitian\\
Potential with $\Theta_0=0$   & {\bf III}  & Hyper-Hermitian     \\
Constant Solution   & {\bf III}    & Hyper-Hermitian                   \\
Non-elementary Function  & {\bf I}  & Hyper-Hermitian       \\
 \bottomrule
\end{tabular}
\caption{ This table sketches the five examples discussed in  Section 4. It contains a classification of the worked metrics, in terms of their Petrov types and whether it corresponds with the hyper-K\"ahler or hyper-Hermitian types.}
\label{tab:ResumeResults2}
\end{table}

\subsection{A Non-elementary Function}
\label{Example4}
Our procedure is not only limited to the case of $\Theta_A$ given by eq.~\eqref{twofunctionstheta}, but also applies to other potentials that are solutions to the eq.~\eqref{igualdad}. In this case, there is a class of solutions of the form ref. \cite{Dunajski:1998nj}:
\begin{equation}
\Theta_0 = a x^l, \ \ \ \ \ \ \ \  \Theta_1 = by^k,
\label{PotentialEx4}
\end{equation}
where $k$ and $l$ are integers, and $a,b$ are complex constants. There is a family of solutions depending on the specific values of the complex coefficients $a$ and $b$, and values for the integer power $k$. Hence, the metric is given by
\begin{equation}
d{s^2}= d{u}d{x}+d{y}d{v}+(-a l x^{l-1}+ b k y^{k-1})d{u}d{v}.
\end{equation}
In particular, we find the double copy only when $l=3$
\begin{equation}
d{s^2}= d{u}d{x}+d{y}d{v}+(3 a   x^{2} + b k y^{k-1}) d{u} d{v}.
\end{equation}
From this metric, we calculate the Weyl spinor curvature and its contraction with four spinors $\xi$'s 
\begin{equation}
\Psi_{ABCD}\xi^A\xi^B\xi^C\xi^D = b k \left(k^2-3 k+2\right) Z y^{k-3}-6 a  Z^3.	
\end{equation}
The factorization of its RHS in terms of monomials yields
\begin{multline}
\Psi_{ABCD}\xi^A\xi^B\xi^C\xi^D =- 6 a   Z\\  \times \left(Z-\sqrt{\frac{b  }{6a} k \left(k^2-3 k+2\right) y^{k-3}}\right) \left(Z+\sqrt{\frac{b }{6a} k \left(k^2-3 k+2\right) y^{k-3} }\right).
\end{multline}
We can rewrite the above expression as
\begin{equation}
\Psi_{ABCD}\xi^A\xi^B\xi^C\xi^D =- 6 a  Z \left(Z-C\right) \left(Z+C\right),
\label{F0}
\end{equation}
where 
\begin{equation}
C \equiv \sqrt{\frac{b  }{6a} k \left(k^2-3 k+2\right) y^{k-3}}.
\end{equation}
The monomials arising from eq. (\ref{WeylPoli}) are given by
\begin{align*}
P_1(Z)&= 1,  & P_2(Z)&=Z, \\
P_3(Z)&= Z-C,     &   P_4(Z)&=Z+C.
\end{align*}
The double copy implies
\begin{align}
\Psi_{ABCD}\xi^A\xi^B\xi^C\xi^D  &= -6 a   Z \left(Z-C\right) \left(Z+C\right)  \nonumber\\
&=  \frac{\Phi^{(1)}_{A B} \Phi^{(2)}_{CD}}{\Phi} 	\xi^A\xi^B\xi^C\xi^D,
\end{align}
where we identify
\begin{equation}
\label{alphaEx4}
\alpha (x^a)=-6 a.
\end{equation}
Then Maxwell field strength spinors are given by
\begin{align}
\Phi^{(1)}_{AB} &= \left(
\begin{array}{cc}
-C& 0 \\
0 &    1
\end{array}
\right) h_1 (x^a), \\
\Phi^{(2)}_{AB} &= \left(
\begin{array}{cc}
0 & \frac{1}{2} \\
\frac{1}{2} &    0
\end{array}
\right) h_2 (x^a).
\end{align}
Maxwell's equations in the vacuum imply
\begin{align}
\label{1Maxwell}
\frac{\partial \Phi^A{}_B}{\partial x^A}=0 ,\\
\label{2Maxwell}
\frac{\partial \Phi^A{}_B}{\partial u^A}=0.
\end{align}
For these equations to hold, it is necessary that
\begin{align}
\label{funciones1}
h_1 (x^a) &= c_1 , \\
\label{funciones2}
h_2 (x^a) &= c_2, 
\end{align}
where $c_1, c_2 \in \mathbb{R}$ are constants and consequently we have
\begin{align}
\Phi^{(1)}_{AB} &= \left(
\begin{array}{cc}
-C& 0 \\
0 &    1
\end{array}
\right) c_1, \label{F1}\\
\Phi^{(2)}_{AB} &= \left(
\begin{array}{cc}
0 & \frac{1}{2} \\
\frac{1}{2} &    0
\end{array}
\right) c_2. 
\label{F2}
\end{align}
\color{black}
Using eqs.~\eqref{funciones1}, \eqref{funciones2} and \eqref{alphaEx4} in eq.~\eqref{relationship}  we get
\begin{equation}
\Phi=c_3,
\label{F3}
\end{equation}
where $c_3 \in \mathbb{R}$ is a constant and $\Phi$ satisfies the wave equation. We do not need to use eq.~\eqref{ZeroCopy} because here $\Phi$ is simpler. We then observe that eqs.~\eqref{F0},~\eqref{F1},~\eqref{F2}, and~\eqref{F3} satisfy eq.~(\ref{WDC2}), which shows that our procedure is possible for other holomorphic functions.


\label{S-FinalRemarks}

\section{Final Remarks}

In the present article, we studied the Weyl double copy for hyper-K\"{a}hler and hyper-Hermitian manifolds. We used local null coordinates, and we showed that there exists a double copy for hyper-Hermitian and hyper-K\"{a}hler manifolds of type \textbf{I}, \textbf{III} and \textbf{N}. These were found for six classes of solutions (see Tables 2 and 3), inspired by the Sparling-Tod. We discussed the double copy for the hyper-K\"ahler case in Sections 4.1 and 4.2, where there is a class of solutions more general, but for certain values of parameters, it is recovered the usual Sparling-Tod double copy. It was confirmed that for the case of  hyper-K\"{a}hler manifolds all our solutions satisfied  Pleba\'nski second heavenly equation. The fields we plotted have asymptotes similar to the Sparling-Tod example. Thus we suggest that this is explained by the fact that our solutions were inspired by Sparling-Tod and Eguchi-Hanson metrics. Specifically, our families of solutions contain more singular points than the previously mentioned ones. One generic feature of the hyper-Hermitian case we found is that in general there are two different Maxwell's field strengths and that one of these fields is associated with a source with a calculable current, which is not conserved (see eq. (\ref{nonconserva}) in subsection (3.3)). This is compatible with the fact that in general the hyper-Hermitian spaces are not Ricci-flat. Regarding this point, it is a known fact that in the single copy of Einstein-Maxwell theory there are two U(1) gauge fields, one of them has a source related to the source current of the ordinary Maxwell field \cite{Easson:2020esh,Easson:2022zoh,Armstrong-Williams:2024bog}. It would be interesting to study if this fact is linked to our result for the hyper-Hermitian single copy with two Abelian gauge fields.  

The next interest would be looking for the existence of double copy for solutions of Petrov type {\bf II}. The origin of double copy suggests the possibility of describing type {\bf II} metric in terms of Maxwell spinors of the single copy. It would be interesting to study these cases or their explanation of their absence to shed light on the double copy correspondence itself and the one of hyper-Hermitian manifolds. Another open problem that we will address in the near future is the twistor description of the double copy for hyper-Hermitian manifolds. Finally, we would like to seek whether there exists a double copy of Kerr-Schild for hyper-Hermitian manifolds without being hyper-K\"ahler.

\vspace{2cm}
\centerline{\bf Acknowledgements} \vspace{.3cm} It is a pleasure to thank A. Luna for very useful comments and suggestions and to M. Dunajski for useful insights and correspondence. G. Robles would
like to thank CONAHCyT for a grant.


\vskip 2truecm 

\end{document}